\documentclass[12pt]{article}
\usepackage{graphicx,amsfonts}
\usepackage{longtable}
\usepackage{amsmath}
\usepackage[all]{xy}

\newcommand{\be}{\begin{equation}}
\newcommand{\ee}{\end{equation}}
\newcommand{\ba}{\begin{eqnarray}}
\newcommand{\ea}{\end{eqnarray}}
\newcommand{\baa}{\begin{eqnarray*}}
\newcommand{\eaa}{\end{eqnarray*}}

\begin{document}

\title{AC conductance and non-symmetrized noise at finite frequency in quantum wires and carbon nanotubes}
\author{In\`es Safi$^a$, Cristina Bena$^{b,a}$, Adeline Cr\'epieux $^c$\\
{\small \it $^a$Laboratoire de Physique des Solides, Universit\'e Paris-Sud},
\vspace{-.1in}\\{\small \it  B\^at.~510, 91405 Orsay, France}\\
{\small \it $^b$Institut de Physique Th\'eorique, CEA/Saclay, CNRS, URA 2306},
\vspace{-.1in}\\{\small \it  Orme des Merisiers, F-91191 Gif-sur-Yvette, France}\\
{\small \it $^c$Centre de Physique Th\'eorique, Universit\'e de la
M\'editerran\'ee,}
\vspace{-.1in}\\{\small \it  163 avenue de Luminy, 13288 Marseille, France}}
\maketitle

\begin{abstract}
We calculate the AC conductance and the finite-frequency non-symmetrized noise in interacting quantum wires and single-wall carbon nanotubes in the presence of an impurity. We observe a strong asymmetry in the frequency spectrum of the non-symmetrized excess noise, even in the presence of the metallic leads. We find that this asymmetry is proportional to the differential excess AC conductance of the system, defined as the difference between the AC differential conductances at finite and zero voltage, and thus disappears for a linear system. In the quantum regime, for temperatures much smaller than the frequency and the applied voltage, we find that the emission noise is exactly equal to the impurity partition noise. For the case of a weak impurity we expand our results for the AC conductance and the noise perturbatively. In particular, if the impurity is located in the middle of the wire or at one of the contacts, our calculations show that the noise exhibits oscillations with respect to frequency, whose period is directly related to the value of the interaction parameter $g$.
\end{abstract}

\maketitle

\section{Introduction}
Electronic transport is an important tool for accessing the physical properties of mesoscopic systems. Besides the average current flowing through a system, information can also be extracted from the fluctuations (noise) in the current.
For example, the zero-frequency noise has been used to prove the fractionalized nature of the quasiparticles in fractional quantum Hall liquids \cite{glattli}. Moreover, the finite-frequency noise contains important information about the statistics \cite{statistics}, as well as about the typical energy scales and the dynamics of excitations \cite{gabelli} in a mesososcopic system. The symmetrized finite-frequency noise corresponds to the Fourier transform of the symmetrized correlator of two non-commuting current operators at two different time points. This noise is even with respect to frequency. Nevertheless, recent experiments have allowed access to the non-symmetrized noise \cite{lesovik,creux}, and thus to the emission and the absorption components of the noise spectrum \cite{deblock,billangeon,glattli1}. What is usually measured in these experiments is the excess non-symmetrized noise, defined as the difference between the non-symmetrized noise at finite voltage and at zero voltage. For non-interacting systems, in the framework of the scattering approach, the total non-symmetrized noise is not even with respect to frequency, but the excess non-symmetrized noise is, hence the emission excess noise and the absorption excess noise are identical.

Few theoretical papers have addressed the effect of interactions on the finite-frequency symmetrized noise, which has been studied for instance for mesoscopic capacitors \cite{blanter_buttiker} and in the Coulomb blockade regime \cite{coulomb_blockade}. It has also been considered in the case of Luttinger liquids (LL), where the interactions are very strong and can give rise to exotic phenomena such as charge fractionalization, spin-charge separation, and fractional statistics. The symmetrized high-frequency noise in chiral LL's (such as fractional quantum Hall effect (FQHE) edge states) has been studied in Ref.~\cite{chamon}; for non-chiral LL's (such as quantum wires and carbon nanotubes connected to metallic leads), it has also been studied in Refs.~\cite{nagaosa,dolcini,recher,thierry}. In those works it was shown that, while the charge fractionalization is still present \cite{safi,kv} and can be extracted from the noise at high frequencies, the presence of the metallic leads obscures it in the zero-frequency noise. It was also found that the interactions play an important role for the entire range of frequencies, even in the zero-frequency limit when the noise decays as a power-law of the applied voltage \cite{noise01,noise0}.

Even fewer theoretical works have dealt with the non-symmetrized finite frequency noise in the presence of interactions. For instance this has been investigated in cotunneling between two quantum dots \cite{eugene}, and for chaotic cavities \cite{hekking}. It has also been analysed for chiral LL's such as FQHE edge states \cite{bena}, where the non-symmetrized excess noise was found to be asymmetric.
The main purpose of this paper is to investigate the finite frequency non-symmetrized noise in {\it non-chiral} Luttinger liquids connected to reservoirs. We show that the asymmetry is determined by the AC differential conductance, which to our knowledge has not been so far investigated for a LL. 

\subsection{AC conductance of a LL connected to metallic leads} 

The AC differential conductance has the advantage that, while containing significant information about the system, it is easier to measure than the high-frequency noise.  However, this conductance has also its drawbacks compared to the finite frequency noise: it can only be defined in the quasi-equilibrium regime when the frequency is smaller then the inverse of the inelastic scattering times $\tau_{in}$ in the reservoirs. This ensures that the time scales one can explore are longer than the time $\tau_{in}$ required for the reservoirs to relax into its quasi-equilibrium state. In quantum wires fabricated using two-dimensional electron gases the transport is coherent if $L\ll v_F\tau_{in}$, and the AC conductance gives information on a regime of relatively small frequencies: $\omega\ll 1/\tau_{in} \ll \omega_L\approx v_F/L$. This limitation of the AC conductance can be relaxed if the reservoir has a sufficiently short $\tau_{in}$ and is of a different material than the one-dimensional wire (for carbon nanotubes). Here we show that its real part can also be obtained from the asymmetry in the non-symmetrized excess noise, without limitation on frequency in this case.

The AC conductance of a clean LL connected to metallic leads has been studied theoretically in Ref.~\cite{safi}. Also, the current in a chiral infinite LL in the presence of a finite AC voltage and a single impurity has been studied in Refs.~\cite{lin_fisher,guigou}.  Experimentally the AC conductance of chiral edges in the integer quantum Hall regime has been studied in Ref.~\cite{gab1}. In this paper, we focus on the non-linear dependence of the AC conductance on the applied DC voltage in the limit of a vanishing AC modulation. We analyze a single-channel interacting wire of length $L$ connected to metallic leads using the inhomogeneous LL model (see Fig.~1). A weak impurity is responsible for the appearance of a backscattering current when a voltage difference is applied between the leads. We analyze the effect of the impurity using the out-of-equilibrium Keldysh formalism \cite{keldysh}. The single impurity scenario may correspond to either a bulk impurity, or to an impurity located at one of the contacts. While in general nanotubes are clean, and most of the backscattering comes from the imperfect contacts, the situation of a single bulk impurity can be achieved experimentally using for example an unbiased scanning tunnel microscope (STM) tip. In this case the effect of the bulk impurity will dominate over the effect of the impurities at the contacts \cite{safi,impurity,balents}. The advantage of a single impurity is that we can disentangle much easier the effects of interactions. Indeed, in systems with two impurities, other effects such as Fabry-Perot interferences come into play and make the analysis much harder \cite{recher,claudia}.
The formalism used here to describe the AC conductance is derived in Ref.~\cite{ines_FDT}. We find that the excess AC conductance, defined as the difference between the AC conductance at finite and zero DC voltage, while being zero for a linear system (in the absence of interactions), has a rich non-linear behavior dominated by impurity effects for an interacting non-chiral LL.

\subsection{Finite frequency non-symmetrized noise}

Besides the AC conductance, we focus also on the analysis of the noise. We start by analyzing the zero-frequency noise, and we note that if the applied voltage is much smaller than the characteristic energy associated with the length of the tube $\omega_L= v_F/g L$ (short-wire limit), both the noise and the current are linear with voltage. Moreover, if the voltage increases above $\omega_L$, the noise displays finite-size features (oscillations with respect to voltage), as well as infinite interacting-wire features (a power-law decay similar to the one mentioned in Refs.~\cite{noise01,noise0}).

Subsequently we analyze the dependence of the non-symmetrized noise on frequency. As mentioned before, the non-symmetrized excess noise was shown to be asymmetric for FQHE edge states \cite{bena}. The main purpose of this paper is to investigate whether the non-symmetrized excess noise is also asymmetric for quantum wires and carbon nanotubes in the presence of the metallic leads, and to identify the origin of this asymmetry.
We find that the excess noise is indeed asymmetric, and
we find that its asymmetry is given by the excess differential AC conductance. In analogy with the excess noise, this is defined as the difference between the differential conductances at finite voltage and at zero voltage. Our observation is consistent with a generalized Kubo formula \cite{ines_FDT,gavish}. Thus, we can trace the asymmetry in the spectrum of the noise to the non-linearity of the system in the presence of interactions.

Moreover, if the impurity is in the middle  or at one end of the wire,  the noise exhibits oscillations whose periodicity is inversely proportional to the value of the fractional charge. The presence of oscillations is the consequence of the quasi-Andreev reflection of an electron at the interface between the interacting quantum wire and the metallic leads \cite{safi,egger_grabert}. The multiple quasi-Andreev reflections give rise to Fabry-Perot type of processes, and to an oscillating behavior of the AC conductance, even in the absence of impurity scattering \cite{safi,hekking_blanter_buttiker}. The existence of the oscillations is a crucial difference between the LL model and an alternative model, the dynamical Coulomb blockade (DCB) \cite{ingold_nazarov}, which was shown to give rise to the same type of power-law I-V decay as the LL theory \cite{safi_saleur}. The presence of the oscillations in the dependence of the noise and AC conductance on frequency, as well as in the dependence of the backscattering current on voltage \cite{dolcini_current}, will be a clear signature of LL physics and will allow one to distinguish between the LL model and the DCB model.

While for a short wire the noise deviates only slightly from the non-interacting limit, when the length of the tube is much larger than the inverse of the applied voltage the signature of the interactions is much more pronounced. In this case the envelope of the oscillations in the noise is given by the form of the non-symmetrized noise for an infinite LL with the same interaction parameter. Also, like for symmetrized noise \cite{dolcini}, the average of the emission noise over the first half-oscillation allows one to extract the value of the fractional charge in the system, in a broader range of experimental conditions than the average of the symmetrized noise.

\vspace{.3in}

This paper is organized as follows: in section 2 we present the model we use to describe the quantum wire connected to metallic leads. In section 3 we present the differential AC conductance of the wire. In section 4, we present the excess non-symmetrized noise, and relate the asymmetry in the noise to the AC conductance.
In section 5 we particularize the results obtained in sections 3 and 4 to the limit of a small impurity, when the AC conductance and the noise can be analyzed perturbatively. In section 6 we discuss our results; in section 6.1 we show that the average of the emission spectrum allows one to obtain the value of the fractional charge, in section 6.2 we present the AC conductance and the non-symmetrized noise on a gate, in section 6.3 we generalize our results for a nanotube that has four channels of conduction, and in section 6.4 we discuss the relevant experimental regimes. We conclude in section 7. The details of the calculation are presented in the Appendices.

\section{Model}

\begin{figure}[htbp]
\vspace{0.3cm}
\begin{center}
\includegraphics[width=8cm]{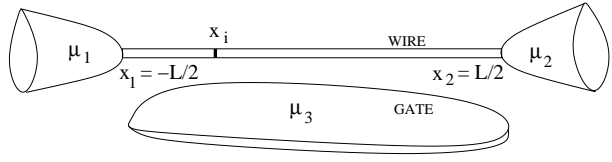}
\vspace{0.15in} \caption{\small  A quantum wire
with an impurity located at position $x_i$, adiabatically coupled to metallic leads and to a metallic gate at chemical potential $\mu_3=eV_3$. The leads are held at different chemical potentials $\mu_1=eV_1$ and $\mu_2=eV_2$.}
\label{setup}
\end{center}
\end{figure}

A quantum wire connected to metallic leads is described by the Hamiltonian
\begin{equation}
{\mathcal{H}} ={\mathcal{H}}_{0}  \, + \, {\mathcal{H}}_{B}  \, +
\, {\mathcal{H}}_{V} \; , \label{L}
\end{equation}
where ${\mathcal{H}}_{0}$ describes the interacting wire and
the leads in the framework of the inhomogeneous LL model, ${\mathcal{H}}_{B}$ describes the effects of the impurity, and
 ${\mathcal{H}}_{V}$ describes the chemical potential applied to the wire.
Explicitly, using the bosonic field $\Phi$ related to the density through $\rho=\partial_x\Phi/\pi$, and letting $\Pi$ be the conjugate field of $\Phi$,
$[\Phi(x),\Pi(y)]= i\delta(x-y)$, one has:
\begin{eqnarray}
{\mathcal{H}}_0 &=&\frac{\hbar v_F}{2}  \int_{-\infty}^{\infty}
 dx \left[ \Pi^2 + \frac{1}{g^2(x)}
(\partial _x\Phi )^2\right]  \, , \label{L0}  \\
{\mathcal{H}}_B &=& \lambda\cos{[\sqrt{4 \pi} \Phi(x_i,t)+2 k_F
x_i]} \label{LB} \; ,\\
{\mathcal{H}}_{V}  &=&    - \int_{-\infty}^{\infty}
\frac{dx}{\sqrt{\pi}} \,\mu(x) \,
\partial_x \Phi(x,t) \; . \label{LV}
\end{eqnarray}
The interaction parameter $g(x)$ is space-dependent and
its value is $g$ in the bulk of the wire, and 1 in the
leads \cite{safi,maslov}. For convenience, the end-points of the wire are denoted by $x_1=-L/2$ and $x_2=L/2$, while the impurity position is chosen to be $x_i$. The backscattering amplitude is denoted by $\lambda$. A schematic view of the system is shown in
Fig.~\ref{setup}.

The function $\mu(x)=eV(x)$ in Eq.~(\ref{LV}) describes the external chemical potential, and is
taken to be piecewise constant \cite{dolcini,safi}:
\begin{equation}
\mu(x) = \left\{
\begin{array}{ll}
\mu_1 & \mbox{for } x < x_1 \\
& \\
\mu_3 & \mbox{for }  x_1<x<x_2\\
& \\
\mu_2 & \mbox{for }  x > x_2 \\
\end{array}
\right. \label{mu-profile}
\end{equation}
where $\mu_3=eV_3$ is controlled by the gate potential, and we will denote $V=V_2-V_1$. The specific profile of the DC electric field can be inferred using $E(x)=-\partial_x V(x)$. Notice that the impurity also contributes in principle to the potential profile, by causing a discontinuous voltage drop at the impurity site due to the coupling between the long-wavelength part of the density and the impurity through the forward scattering term, however for a static impurity this should not affect our results. Also the local effective electrostatic potential is modified by the backscattering of quasiparticles at the impurity site (see e.g. \cite{fisher}), but only the external potential is relevant for the quantities of interest in our analysis. 

In bosonization, the current operator is related to the bosonic
field $\Phi$ through
\begin{equation}
j (x,t) = \frac{e}{\sqrt{\pi}} \partial_t
\Phi(x,t) \; . \label{current} \\
\end{equation}
In our analysis we will focus mainly on the currents evaluated at the contacts $x_1,x_2$, while adopting the convention that outgoing currents are positive. Thus we denote $$j_n(t)=(-1)^n j(x_n,t),$$ for $n=1,2$, and $I_n(t)=\langle j_n(t) \rangle$.

The differential AC conductance of the wire is defined as the response of the system to an infinitesimal AC modulation in the bias of the reservoirs: $V_m\rightarrow V_m(t)=V_m+{\cal V}_m(t)$, with ${\cal V}_m(t)=v_m \cos \omega t$. Thus the AC conductance $G_{nm}(\omega)=\int dt e^{i \omega t} G_{nm}(t)$ is defined as the Fourier transform of the functional derivative $G_{nm}(t)$, where:
\begin{equation}\label{conductance_definition}
G_{nm}(t-t')=\frac {\delta I_n(t)}{\delta {\cal V}_m(t')}\bigg|_{{\cal V}_m=0}.
\end{equation}

As we show here, at low temperatures the AC conductance $G_{nm}$ of a LL has a non-linear dependence on the applied DC chemical potentials.

The AC conductance of the wire can indeed be related to the non-local AC conductivity \cite{safi}. To show this, one notes that the external time-dependent
electric field profile corresponding to the piecewise potential landscape in Eq.~(\ref{mu-profile}) is given by:
\begin{equation}\label{E}
E(x,t)=\sum_{m=1,2}(-1)^m[V_m(t)-V_3(t)]\delta(x-x_m).
\end{equation}

The non-local differential AC conductivity is defined as the linear response of the current to an infinitesimal AC modulation of the electric field, $E(x)\rightarrow E(x)+{\cal E}(x,t)$, where ${\cal E}(x,t)={\epsilon}(x)\cos \omega t$, at a finite value of $E$:
\be
\sigma(x,y,t-t')=\frac{\delta \langle j(x,t) \rangle}{\delta {\cal E}(y,t')}\bigg|_{{\cal E}=0}.
\label{sig}
\ee
Correspondingly we have $\sigma(x,y,\omega)=\int dt e^{i \omega t} \sigma(x,y,t)$.
We can thus see that one can express the AC conductance in Eq.~(\ref{conductance_definition}) as:
\begin{equation}
G_{nm}(\omega)=(-1)^{n+m}\sigma(x_n,x_m,\omega).
\end{equation}

We should note that in the case of a time dependent current flowing through the system, the conservation of current does not hold, i.e. $I_{1}(t) + I_{2}(t) \ne 0$, and a time-dependent charge accumulates on the wire. This induces similar fluctuations of the charge on a gate capacitatively  coupled to the wire, the current on the gate is equal to $I_1+I_2$, thus ensuring formally current conservation. However, besides its role in screening the Coulomb interactions in the wire, the presence of the gate has no direct effect on the values of the various currents flowing through the wire. The gate will be discussed in more detail in section 6.2.

The other quantity of interest of this analysis is the finite frequency non-symmetrized noise. This is defined as
\begin{eqnarray}
S_{nm}(\omega) =\int_{-\infty}^{\infty} dt e^{i\omega t}
 \left\langle \delta j_m(0)\delta j_n(t)
\right\rangle ,
\end{eqnarray}
where $n$ and $m$ refer to the reservoirs where the current is measured and $\delta j_n(t)=j_n(t)-\langle j_n\rangle$.

The finite frequency symmetrized noise on the other hand is defined as
\begin{eqnarray}
S^+_{nm}(\omega) =\frac{1}{2}\int_{-\infty}^{\infty} dt e^{i\omega t}
 [\left\langle\delta j_m(0)\delta j_n(t)
\right\rangle
+\left\langle\delta j_n(t)\delta j_m(0)
\right\rangle]  .
\label{sym}
\end{eqnarray}

\section{Differential AC conductance}

We will first focus on the  AC conductance of the wire.
As shown above we can relate the AC conductance of the wire with the non-local AC conductivity $\sigma(x,y,\omega)$ evaluated between specific values of $x$ and $y$. In turn,
the AC non-local conductivity $\sigma(x,y,\omega)$, defined in Eq.~(\ref{sig}) needs to be expressed in terms of microscopic correlators. For this purpose, in the equilibrium case, when $e V\ll k_B T$, one can use simply the Kubo formula, and $\sigma(x,y,\omega)$ coincides with the non-local conductivity discussed already in Refs.~\cite{dolcini,safi,bena}.
In this regime a Dyson-type equation was derived:
\begin{equation}\label{Dyson}
\sigma(x,y,\omega)=\sigma^0(x,y,\omega)-\frac{h^2}{e^4}\sigma^0(x,x_i,\omega)
G_B(\omega)\sigma^0(x_i,y,\omega).
\end{equation}
We should note that $\sigma^0$, the non-local conductivity without impurity,  describes the propagation from the measuring point to the impurity point, while $G_B$ describes the pure backscattering conductivity at the impurity position $x_i$. Notice that $\sigma^0(x,y,\omega)$ does not depend on the voltage $V$ as the system is purely linear in the absence of an impurity, but depends only on the frequency $\omega$, and on the scale $\omega_L=v_F/gL$ associated with the finite size of the wire. The precise form of $\sigma^0$ for $x$ and $y$ in the wire or at the contacts (i.e., $|x|,|y|\le L/2$) has been calculated previously \cite{safi}:
\be
\sigma^0(x,y,\omega)=g\frac{e^2}{h}\left[e^{i\frac{\omega}{\omega_L}
\left|\frac{x-y}{L}\right|}+\frac{\gamma}{e^{-2i\omega/\omega_L}-\gamma^2}
\sum_{r=\pm}\left(\gamma e^{ir\frac{\omega}{\omega_L}\left|\frac{x-y}{L}\right|}
+e^{i\frac{\omega}{\omega_L}\left(r\frac{x+y}{L}-1\right)}\right)\right]~,
\ee
where $\gamma=(1-g)/(1+g)$ is the reflection coefficient for the quasi-Andreev reflection at the contacts, and $L$ is the length of the wire. We should note that at zero frequency $\sigma^0(x,y,\omega)=e^2/h$, independent of position, and equal to the conductance of non-interacting single-channel one-dimensional system.

On the other hand, $G_B$, besides the impurity position $x_i$ and temperature $T$, depends also on the backscattering amplitude $\lambda$. Also, while it does not depend on voltage in the linear regime $e V \ll {\rm max} (k_B T,\hbar\omega)$, it will depend on it in the non-linear regime, $e V > {\rm max}({k_B T,\hbar\omega})$. It is given by:
\begin{equation}
G_B(\omega) =\frac{1}{\hbar \omega} \int_0^\infty dt \left( e^{i \omega t}-1 \right)
\left\langle \left[ j_B(t),j_B(0) \right]
\right\rangle .
\label{sb}
\end{equation}

Here, the backscattering current operator is defined as
 \begin{equation}
j_{B}(t) =  -\frac{e }{\hbar \sqrt{4 \pi}}\frac{\delta
\mathcal{H}_B(\Phi,t)}{\delta \Phi(x_i,t)} =\lambda \frac{e}{\hbar}\sin[\sqrt{4 \pi} \Phi(x_i,t)+2 k_F
x_i+eVt/\hbar] \label{jb1_def} .
\end{equation}
This form for the backscattered current operator was obtained using a time-dependent translation of $\Phi$ in $\mathcal{H}_B$ incorporating the effect of the applied voltage $V$ \cite{dolcini}.
Its average value is denoted by $I_B(t)=\left<j_B(t)\right>$.
We should stress that it is important to distinguish between the total current operator defined in Eq.~(\ref{current}), and the backscattered current defined above. The relation between the average values of these currents becomes simple in the DC regime, when their values are time-independent: $I_2=-I_1=V e^2/h-I_B$, as sketched in the linear regime in Refs.~\cite{safi,impurity}, and in the nonlinear regime in Refs.~\cite{dolcini,dolcini_current}. However, the time-dependent properties associated with these two current operators, such as the AC conductivities and noise spectra are different.

On the other hand, for the non-equilibrium case $e V\gg k_B T$, it turns out that a general out-of-equilibrium Kubo formula allows one to relate the AC conductivity to the retarded current-current correlation function, even in the presence of a finite DC bias. This was proved for the case of homogeneous conductivity \cite{gavish} with the requirement of a stationary density matrix. This misses however the effects of the non-locality, which are important in a mesoscopic context. A simpler demonstration, not constrained by the stationarity requirement, and valid more generally for any finite mesoscopic non-linear system, is presented in Ref.~\cite{ines_FDT}.  Thus the AC conductivity is shown to verify the Dyson equation presented in Eq.~(\ref{Dyson}), with the sole difference that in this case $G_B$ has an implicit dependence on the voltage $V$. This implies that the AC conductance $G_{mn}$ defined in Eq.~(\ref{conductance_definition}), for a quantum wire in the presence of an impurity can be obtained directly using Eq.~(\ref{Dyson}). While an analytic calculation of $G_B$ cannot be done for all impurity strengths, the conductance can be calculated perturbatively for the case of  a small impurity. The perturbative analysis of the AC conductance is presented in section 5.



\section{Non-symmetrized noise}
We will now present our results for the noise in a quantum wire in the presence of an impurity, and connect it to the AC conductance. We will discuss some general considerations for the noise, which are independent of the strength of the impurity potential. While it is important to understand these general aspects, same as for the AC conductance, the detailed form of the noise dependence on frequency cannot be obtained exactly, but only by using a perturbative expansion in the limits when the impurity is very small or very large respectively. The perturbative analysis of the very small impurity situation will be presented in detail in the next section.

Notice first that for an applied voltage $V$ much smaller than temperature (in equilibrium) the noise is given by:
\ba\label{equilibrium}
S^0_{nm}(\omega)&=&2\hbar \omega N(\omega) {\rm Re}[G_{nm}(\omega)]_{V=0},\nonumber\\
S^{0+}_{nm}(\omega)&=&\hbar \omega [1+2 N(\omega)] {\rm Re}[G_{nm}(\omega)]_{V=0},
\ea
where $N(\omega)=\left[\coth\left(\hbar\omega/2k_BT\right)-1\right]/2$. This is in agreement with the fluctuation-dissipation theorem (FDT).

For arbitrary voltages, temperatures and frequencies, as well as for any impurity strength, we find the total noise to be:
\begin{eqnarray}\label{Snoise}
S_{nm}(\omega)&=&2\hbar \omega N(\omega) {\rm Re} [G_{nm}^0(\omega)]+\frac{h^2}{e^4}G^0_{ni}(\omega)S_B(\omega)G_{im}^0(-\omega)
\nonumber\\&&
-2\frac{h^2}{e^4}\hbar \omega N(\omega)  \Big\{{\rm Re}[G^0_{ni}(\omega)]G_B(-\omega)G^0_{im}(-\omega)
+G^0_{ni}(\omega)G_B(\omega){\rm Re}[G^0_{im}(-\omega)]
\Big\}~,\nonumber\\
\end{eqnarray}
where $G_B$ was defined in the previous section and $S_B$ is the non-symmetrized backscattering noise
\begin{eqnarray}
S_B(\omega)=\int_{-\infty}^{\infty}\left<\delta j_B(0)\delta j_B(t)\right>e^{i\omega t} dt,
\label{sb1}
\end{eqnarray}
where $\delta j_B(t)=j_B(t)-\langle j_B \rangle$.

We should mention a non-trivial check satisfied by Eq.~(\ref{Snoise}), which has more general and important consequences: the noise verifies a generalized  out-of-equilibrium Kubo-type relation \cite{ines_FDT},
\begin{equation}\label{fdt}
S^{-}_{nm}(\omega)\equiv S_{nm}(\omega)-S_{nm}(-\omega)=-2\hbar \omega {\rm Re}[G_{nm}(\omega)].
\end{equation}

The first consequence of this relation is that if one has access to the emission noise, one can deduce the absorption noise by using the AC conductance discussed in the previous section. Reversely, if one measures the emission and the absorption noises, one can extract the AC conductance at an arbitrary frequency, not necessarily smaller than the inverse of the inelastic scattering time in the reservoirs. This provides an advantage over  the case of a direct AC measurement \cite{safi}.

The second consequence is that we can write the noise as a combination of a symmetric $S_{nm}^+(\omega)$ and anti-symmetric part $S_{nm}^-(\omega)$, where the symmetric component was defined in Eq.~(\ref{sym}),
and can be related to the total noise by
$S_{nm}^+(\omega)=[S_{nm}(\omega)+S_{nm}(-\omega)]/2$. Thus the difference between the symmetrized noise (computed in Ref.~\cite{dolcini}), and the non-symmetrized noise comes from the real part of the  AC conductance which is explored here for the first time for the case of a LL:
\begin{eqnarray}\label{NSnoise}
S_{nm}(\omega)=S^+_{nm}(\omega)-\hbar\omega{\rm Re}[G_{nm}(\omega)]~.
\end{eqnarray}

The third consequence of this out-of equilibrium FDT relation is that the excess noise, defined as the difference between the noise at finite bias and the noise at $V=0$, while symmetric for a linear system, becomes asymmetric for a non-linear interacting system \cite{ines_FDT}.
We find the excess noise to be given by:
\begin{eqnarray}
\Delta S_{nm}(\omega)&=&\Delta S_{nm}^+(\omega)-\hbar\omega Re[\Delta G_{nm}(\omega)]~,
\end{eqnarray}
where $\Delta G_{nm}(\omega)=G_{nm}(\omega)-G_{nm}(\omega)|_{V=0}$ is the excess AC conductance, and $\Delta S_{nm}^+(\omega)$ is the symmetrized excess noise. For a linear system, the AC conductance is independent of voltage, thus $\Delta G_{nm}(\omega)=0$, and the non-symmetrized excess noise is equal to the symmetrized excess noise and is therefore even in frequency. However, if the AC conductance is voltage dependent, which  is the case if the system is non-linear, $\Delta G_{nm}$ is non-zero and the non-symmetrized excess noise is non-symmetric.


It is important to note two other properties of our results that hold exactly in the quantum regime $\hbar \omega \gg k_B T$, for any impurity strength. For positive frequencies the factor $N(\omega)$ vanishes on the right-hand side of Eq.~(\ref{Snoise}). Thus the emission noise (i.e., the positive frequency component of the noise) is equal in this regime to the emission excess noise (no equilibrium component for the emission noise). Second, Eq.~(\ref{Snoise}) can be simplified in this frequency range: the emission noise coincides with the backscattering noise $S_B$, up to factors of the non-local pure conductivity:
\begin{equation}
S_{nm}(\omega>0)=\left(\frac {h}{e^2}\right)^2 G^0_{ni}(\omega) S_B(\omega) G^0_{im}(-\omega)~.
\end{equation}
Thus the emission noise has access directly to the impurity backscattering noise. This is a great advantage with respect to the symmetrized noise, in particular for the case a non-chiral system for which  the backscattering noise cannot be simply inferred from the chiral current correlations, as it is the case for a chiral system (e.g the edges of a fractional quantum Hall liquid) presented in Ref.~\cite{bena}.


\section{Perturbative results}
Up to this point we have presented exact formal expressions for the AC conductance and finite frequency noise, for arbitrary temperature and voltage, and for an arbitrary location and strength of the backscattering center. We can easily expand these expressions perturbatively in the weak backscattering regime. From an experimental perspective, this is the situation which is most relevant, since it is possible to fabricate ballistic quantum wires and carbon nanotubes for which weak backscattering is due either to imperfect contacts, or can be induced by an STM tip. From a theoretical perspective, the noise at zero frequency in a chiral LL has been calculated exactly using Bethe ansatz for an arbitrary strength impurity, and it was shown that the perturbative analysis describes within a few percent accuracy the noise for barriers with transmission larger than 50\%, up to very low voltages \cite{noise01}. In general, the validity of perturbation theory requires the existence of an energy scale which cuts the RG flow for the effective backscattering amplitude at a not too large value. This energy scale is usually taken to be the voltage or the temperature. However, if the length  of the wire is short enough, the energy scale associated with the length of the wire can play this role, and one can study the system up to low enough values of the applied voltage. 

One can also analyze the strong backscattering regime perturbatively, as it has been done in Ref. \cite{bena} for chiral LL's, but we do not focus on this regime here.  We should also note that our results are valid for an arbitrary position of the impurity, and even for an arbitrary extended disorder configuration. Nevertheless,
here we restrict ourselves to a localized impurity at the center of the wire (i.e., $x_i=0$).  The analysis of an arbitrary realization of disorder, as well as of the arbitrary strength of the impurity, require a different analysis which is beyond the scope of this work.

\subsection{AC conductance}
\begin{figure}[htbp]
\begin{center}
\includegraphics[width=8cm]{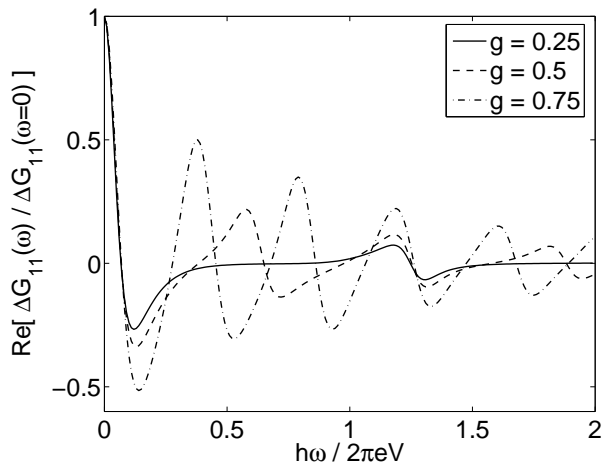}
\vspace{0.15in} \caption{\small Real parts of the excess AC conductance $\Delta G_{11}(\omega)$
in units of $\Delta G_{11}(\omega=0)=-(\partial I_B/\partial V-\partial I_B/\partial V|_{V=0})$ for $g=0.25$ (full line), $g=0.5$ (dashed line) and $g=0.75$ (dash-dotted line). The other parameters are $x_i=0$, $T=0$, $\lambda/eV=0.01$ and $g\hbar\omega_L/eV=0.05$. The excess AC conductance for $g=1$ is zero (not shown), consistent with the fact that the excess noise is symmetric in this case.}\label{conductance}
\end{center}
\end{figure}

We now evaluate the differential AC conductance and the non-symmetrized noise perturbatively in the case of small impurity backscattering $\lambda$, up to order $\lambda^2$.
The real part of the excess AC conductance, $\Delta G_{11}(\omega)=G_{11}(\omega)-G_{11}(\omega)|_{V=0}$ is plotted in Fig.~\ref{conductance}. The advantage of analyzing the excess conductance is that we have access directly to the impurity-generated terms proportional to $\lambda^2$. We do not give here the conductance in the linear regime $G_{11}(\omega)|_{V\ll\omega}$. The inconvenience of analyzing it is that it contains both impurity-induced terms of order $\lambda^2$, and terms that are independent of the impurity (of order $\lambda^0$), that have been already studied in Refs.~\cite{safi,hekking_blanter_buttiker}, and that will dominate over the impurity-induced terms.

The excess AC conductance vanishes for $g=1$, consistent with the linearity of a non-interacting system. However, it is non-zero in the presence of interactions, consistent with the strong non-linearity of an interacting system in the presence of an impurity.

While not depicted here, we also find that the real part of the total diagonal conductance $G_{nn}$ is a positive quantity at all frequencies. Consequently, from Eq.~(\ref{NSnoise}) we expect that
the non-symmetrized noise $S_{nn}$ will be reduced with respect to the symmetrized noise $S^{+}_{nn}$ for positive frequencies (emission part), whereas at negative frequencies (absorption part) it is increased. We begin our analysis with the zero-frequency limit, and consequently analyze the dependence of the noise on frequency.

\subsection{Zero-frequency noise}

Measurements of the zero-frequency noise have been available for quite some time in FQHE edge states \cite{glattli}. Moreover, they have recently been performed also for nanotubes \cite{noise0,kontos}. We calculate the zero-frequency noise perturbatively, up to order $\lambda^2$, and we find:
\begin{eqnarray}\label{zerofreqnoise}
S_{nm}(\omega=0)=e I_B \coth\left(\frac{eV}{2k_BT}\right)+2k_BT\left[\frac{e^2}{h}-2\frac{\partial I_B}{\partial V}\right]~,
\end{eqnarray}
in agreement with \cite{nagaosa,noise0}.
This formula looks like the zero-frequency noise in a non-interacting wire \cite{noise00}, but interaction effects are present in a non-linear dependence of $I_B$ with the applied voltage \cite{noise01}. In the limit of $k_BT\ll eV$, the zero-frequency excess noise is simply given by the electron charge multiplied by the backscattering current: $e I_B$ \cite{schottky}.

\begin{figure}[htbp]
\begin{center}
\includegraphics[width=7cm]{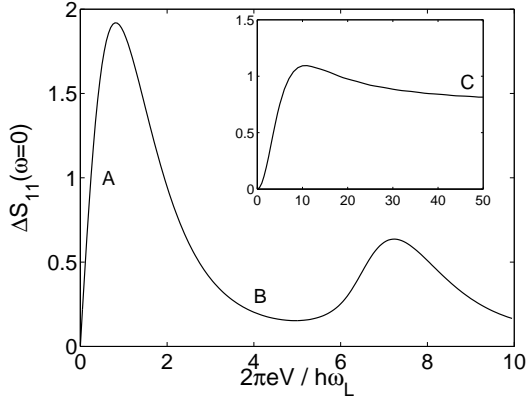}
\vspace{0.15in} \caption{\small The zero-frequency excess noise $\Delta S_{11}$ as a function of applied voltage for $g=0.25$ at $k_B T/\hbar \omega_L=0$ (main plot), and $k_B T/\hbar \omega_L=5$ (inset). The other parameters are $x_i=0$ and $\lambda/\hbar \omega_L=0.01$. The $A$, $B$, and $C$ denote three regimes of interest. The noise is renormalized by $\lambda^2/(\omega_L/\omega_c)^{2g}$, where $\omega_c$ is a high-energy cutoff.}\label{zero_frequency}
\end{center}
\end{figure}

The evaluation of the backscattered current is presented in Appendix C, and discussed in detail in Refs.~\cite{dolcini,dolcini_current}.  In Fig.~\ref{zero_frequency}, we plot the zero-frequency excess noise $\Delta S_{nm}(\omega=0)=S_{nm}(\omega=0)-S_{nm}(\omega=0)|_{V=0}$ as a function of voltage. At zero temperature, we observe periodic modulations of the noise which are attenuated when the temperature increases. Besides, for voltages smaller than $\omega_L$ (in the short-wire limit), both the noise and the backscattered current increase linearly with voltage. Notice that this is in qualitative agreement with the experimental measurements \cite{kontos}, since we expect such behavior even for more complicated impurity distribution.  This regime is denoted in Fig.~\ref{zero_frequency} by $A$. We expect that in this regime the frequency dependence of the noise to be also similar to that of a non-interacting system.
The linear dependence of the current in this regime can be argued using Eq.~(\ref{ib}) in Appendix C. The integral in Eq.~(\ref{ib})  is dominated by times smaller, and of the order of a few $1/\omega_L$. If $e V\ll \hbar \omega_L$, $ e V t/\hbar \ll 1$ and  $\sin(e V t/\hbar)$ can be expanded linearly in $V$, thus justifying the linear dependence.

For voltages larger than $\hbar \omega_L$ we see that the system approaches the infinite-wire limit, while exhibiting finite-size oscillations whose envelope follows the infinite-wire characteristic power-law dependence \cite{noise01}. In Fig.~\ref{zero_frequency}, we denote this regime by $B$.
We expect that in this limit the finite-frequency noise exhibits finite-size features overlapped with infinite wire characteristics.

For temperatures larger than $\hbar \omega_L$, the finite-size features disappear, and we see that the behavior of the noise resembles the noise of an infinite interacting wire: for voltages larger than the temperature the noise decreases as a power-law with respect to the applied voltage (regime $C$). A similar power-law decay of the zero-frequency noise at large voltages was predicted and observed experimentally in Ref.~\cite{noise01}.

\subsection{Finite frequency non-symmetrized noise}
In the following we analyze the non-symmetrized excess noise at finite frequency.  As discussed in the previous section, we focus mainly on the excess noise, as being both the quantity most relevant in an experiment, the one that incorporates most of the information about the electronic interactions in a system, and the one that is directly proportional to the effects of the impurity. Our most important observation is that the non-symmetrized excess noise, while being symmetric in a non-interacting system, becomes asymmetric in the presence of interactions, the amount of asymmetry providing an insight into the strength of the electron-electron interactions. A similar behavior was obtained in a two-dimensional electron gas in the fractional quantum Hall regime \cite{bena}.

We study two relevant limits corresponding to the $A$ and $B$ regimes described in the previous section: a very short tube, when we expect the physics to be dominated by the non-interacting metallic leads, and a very long tube, when we should be able to retrieve some of the infinite Luttinger liquid features. We restrict ourselves to the regime where the temperature is much smaller than all the other energy scales in the problem, even though we could as well include arbitrary temperature, which is however not necessary if one is interested by the quantum regime. \vspace{.1in}
\\
{\bf  A. Short-wire limit}
\vspace{.1in}

\begin{figure}[htbp]
\begin{center}
\includegraphics[width=8cm]{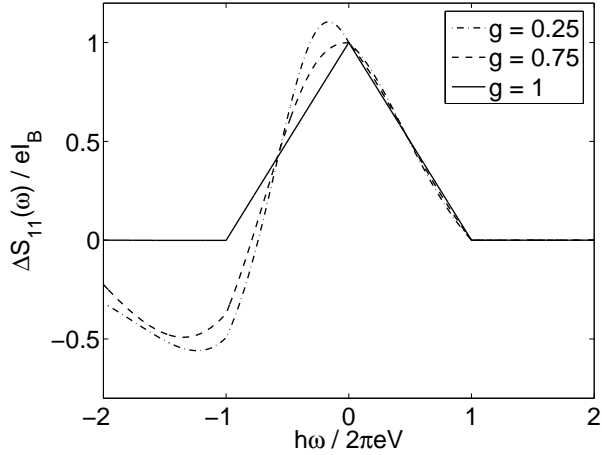}
\vspace{0.15in} \caption{\small Non-symmetrized excess noise $\Delta S_{11}$ divided by
$eI_B$ for different values of the interactions parameter $g$ and for $x_i=0$, $T=0$, $\lambda/eV=0.01$ and $g\hbar\omega_L/eV=1$.}\label{excessnoise1}
\end{center}
\end{figure}

In the first ($A$) case, when $\hbar\omega_L=\hbar v_F/gL \gg e V$, the non-symmetrized excess noise deviates from the non-interacting limit, as can be seen from Fig.~\ref{excessnoise1}, but the deviations, especially for the case of the emission noise, are small. A similar behavior was obtained for the symmetrized noise, either in the same geometry \cite{dolcini}, or for a short carbon nanotube weakly coupled to a STM tip \cite{guigou}.

We also see that, at zero temperature, the non-symmetrized excess noise cancels at positive frequencies for $\hbar\omega > eV$, for all values of $g$.
However, the non-symmetrized excess noise cancels for negative frequencies $\hbar\omega < -eV$
only when $g = 1$, i.e. in the non-interacting limit. This is because the detection of the current fluctuations at positive frequencies requires the
emission of photons, while at negative frequencies it requires the absorption of photons.  Thus for energies larger than $e V$, the emission noise for a non-interacting system vanishes as an electron coming from the source does not dispose of the corresponding empty states in the drain to emit a photon \cite{gavish,cons}. In the presence of interactions, the problem is more complicated, and the symmetrized noise analyzed in Refs.~\cite{chamon,dolcini}, does not allow to draw any conclusion on the issue. 
For the non-symmetrized noise in chiral LL's such as the FQHE edges \cite{bena}, the emission noise vanishes at frequencies higher than $g e V/\hbar$ (probably  due to the lowest-order perturbative nature of the calculation, which takes into account only single quasiparticle processes). On the other hand, here we find that for a non-chiral LL, even in the presence of interactions, the emission noise vanishes for frequencies larger than $e V/\hbar$ (the absorption noise however does not vanish for frequencies smaller than $-e V/\hbar$ due to the contribution of the AC conductance (see Eq.~(\ref{fdt})).

The deviation from the non-interacting limit decreases with decreasing the length of the tube, or with increasing $g$. This is due to finite size effects which dominate in the case of a short wire, and in the extreme limit we expect the system to behave like an infinite non-interacting wire. While for the values presented in Fig.~\ref{excessnoise1}, the difference is not substantial for the emission component, it signals already that even in the presence of the metallic leads, the non-symmetrized excess noise becomes asymmetric due to the effect of the interactions in the wire.

One should note the emergence of the regions where the non-symmetrized excess noise becomes negative. This is contrary to the original intuition that the noise increases when a DC voltage is applied. However, a negative symmetrized excess noise has already been noted for the case of LL's \cite{dolcini}, or for semi-classical systems \cite{Leso-Loos,Bula-Rubi}. Here we see that the emission noise $S_{11}$ remains positive, in agreement with the intuitive understanding. This result is obtained perturbatively, but we believe that it will remain valid at all orders in perturbation theory\cite{note}. On the other hand, we see that the absorption noise can become negative, and we can understand this as stemming from the generalized Kubo formula \cite{ines_FDT,gavish}, which relates the difference between the emission noise and the absorption noise to the AC conductance. The negativity of the absorption noise will yield regions where the symmetrized excess noise can also become negative. \vspace{.1in}
\\
{\bf B. Long-wire limit}
\vspace{.1in}

\begin{figure}[htbp]
\begin{center}
\includegraphics[width=8cm]{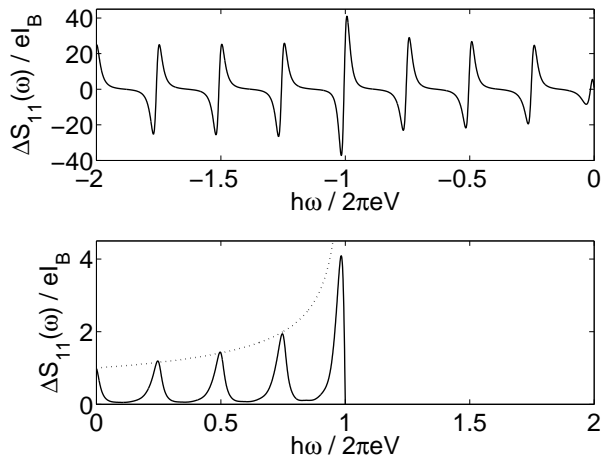}
\vspace{0.15in} \caption{\small (Upper graph) Absorption part and, (lower graph) emission part of the non-symmetrized excess noise $\Delta S_{11}$ divided by $eI_B$ for $g=0.25$, $x_i=0$, $T=0$,
$\lambda/eV=0.01$ and $g\hbar\omega_L/eV=0.01$. The dotted line describes the emission noise of an infinite system with $g=0.25$.}\label{excessnoise2}
\end{center}
\end{figure}

The effects become much more pronounced in the opposite ($B$) limit, $\hbar\omega_L=\hbar v_F/gL \ll e V$, (the long-tube limit).
In this case a large number of oscillations can be observed (see Fig.~\ref{excessnoise2}) for frequencies inferior to the Josephson frequency $e V/\hbar$. The period of these oscillations  is given by $2\pi \omega_L $, and they arise from the quasi-Andreev processes discussed in the Introduction. It is clear from Fig.~\ref{excessnoise2} that the amount of asymmetry between the excess emission noise and the excess absorption in this situation is very large. While the excess emission noise goes to zero at frequencies larger than the Josephson frequency, the excess absorption noise displays sharp oscillations for frequencies smaller than $-eV/\hbar$. Also, the magnitude of the oscillations, even at frequencies larger than $-e V/\hbar$ is much larger for the absorption component.

Other signatures can be extracted from the oscillations of the noise with respect to frequency. For example if the impurity is in the center of the wire, the period of the oscillations $2 \pi v_F/gL$ is inversely proportional to the value of the fractional charge $g$. Also, in the long-tube limit, as depicted in Fig.~\ref{excessnoise2}, the envelope of the oscillations coincides exactly with the noise of an infinite LL with the same interaction parameter $g$ (the dotted line).  We should note that these results are strongly affected by the position of the impurity, such that, if the impurity is not exactly in the middle, the dependence of the noise on frequency is more complicated, and the shape of the envelope changes. Nevertheless, we have checked that if the impurity is at one of the contacts, the periodicity of the oscillations still holds but doubles, and the envelope of the oscillations corresponds to an infinite Luttinger liquid with an effective $g_c=2 g/(g+1)$.\cite{safi}
However, as discussed in the introduction, the position of the impurity should be controllable using an STM tip, and the bulk impurity should dominate over the contact ones \cite{safi,impurity}.

We should note that, in agreement with previous studies \cite{dolcini}, the position of the Josephson singularity is at $e V/\hbar$ at zero temperature, and the form itself of the singularity is cusp-like, same as for the non-interacting system. This can be seen analytically by taking the limit $|e V/\hbar-\omega| \ll \omega_L$. The integrals responsible for the noise in this limit are dominated by times smaller or of the order of a few $1/\omega_L$, for which the oscillatory terms of the form $\sin(\omega t-e V t/\hbar)$ become linear in $(\omega-e V/\hbar)$, hence the cusp singularity is at $\omega=eV/\hbar$.


\section{Discussion}
\subsection{Average non-symmetrized noise}
\begin{figure}[htbp]
\begin{center}
\includegraphics[width=8cm]{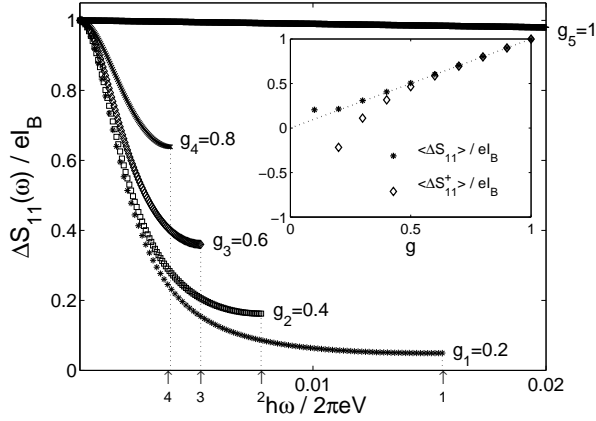}
\vspace{0.15in} \caption{\small Emission excess noise $\Delta S_{11}(\omega)$ divided by
$eI_B$ plotted over the first half period (see arrows) of oscillations (except for $g_5=1$) for different values of the interaction parameters $g$ and for $x_i=0$, $T=0$, $\lambda/eV=0.01$ and $g\hbar\omega_L/eV=0.001$. The inset shows the average of the emission noise over one period ($*$ points) and of the symmetrized excess noise ($\diamond$ points).}\label{excessnoiseaverage}
\end{center}
\end{figure}

Here we show that direct access to the value of the charge fractionalization can be obtained from the finite-frequency emission noise in the long-wire limit. Thus along the lines of Refs.~\cite{dolcini,bena}, we can analyze the average of the non-symmetrized excess noise over the first half period of oscillations when the impurity is in the middle of the wire:
\begin{eqnarray}
\langle \Delta S_{nm}\rangle_{\Delta\omega} =\frac{1}{\Delta\omega}\int_{0}^{\Delta\omega} d\omega \Delta S_{nm}(\omega) \;,
\end{eqnarray}
where $\Delta\omega=\pi\omega_L=\pi v_F/gL$. The period of oscillations depends on the interaction parameter $g$ as depicted in Fig.~\ref{excessnoiseaverage}. While the zero frequency noise is given by $e I_B$, we find that the average of the emission noise over the first half period of oscillations is $g e I_B$ in the regime $eV\gg\{k_BT,\hbar\omega_L\}$ (see inset of Fig.~\ref{excessnoiseaverage}). This is less restrictive than the average
of the symmetrized noise presented in Ref.~\cite{dolcini}.
A measurement of the noise over one half period of oscillations should thus make one able to extract the value of the fractional charge in the interacting wire. This should be easier to achieve experimentally than the measurement of the envelope of the oscillations, as the noise frequencies required are much smaller.

As can be seen from Fig.~\ref{excessnoiseaverage} the average of the emission noise is more accurate than the average of the symmetrized excess noise, thus allowing the identification of the value of the fractional charge for a larger region in parameter space.

We should also note that, if the impurity is not exactly in the middle of the wire, neither at the contact, the frequency average is not strictly equal to the value of the fractional charge, but depends on the impurity position. A similar dependence was observed also for the average of the symmetrized noise calculated in Ref.~\cite{dolcini}. In order to be able to use the present formalism to describe in detecting the value of the fractional charge, one can use an STM tip to make an impurity in the center of the wire that will dominate the scattering. The situation in which the two contacts between the wire and the metallic lead are the main source of backscattering will be examined in a separate work. 

\subsection{Non-symmetrized noise on a gate}
As mentioned in section 2, in the case of an AC current flowing trough the system, the conservation of current does not simply hold in the usual form for DC transport $I_1+I_2=0$. For time-dependent transport, the continuity equation $\partial_x I=0$ must be replaced by $\partial_t \rho+\partial_x I=0$, and we have \cite{safi}:
\begin{equation}
I_1(t)+I_2(t)=-\int_{-L/2}^{L/2} dx \left<\partial_t \rho(x,t)\right>=-\dot{Q}(t),
\end{equation}
such that the sum between the currents at the two contacts is related to $Q$, the charge accumulated inside the wire. This charge can be measured using a nearby gate capacitatively coupled to the wire, such that the charge on the gate is equal and opposite to $Q$. The current flowing trough the gate is thus $I_3(t)=-\dot{Q}(t)=-I_1(t)-I_2(t)$, ensuring formally current conservation.

Along the same lines with Eq.~(\ref{conductance_definition}), we can define a gate AC conductance $G_{3m}(\omega)$ as  $G_{3m}(\omega)=\int dt e^{i \omega t} G_{3m}(t) $ where
\be
G_{3m}(t-t')=\frac{\delta I_3(t)}{\delta {\cal V}_m(t')}\bigg|_{v_m=0},
\ee
for $m=1,2$. Thus $G_{3m}(\omega)=-G_{1m}(\omega)-G_{2m}(\omega)$, while
the total gate conductance is defined as
$$G_{33}(\omega) \equiv -G_{31}(\omega)-G_{32}(\omega).$$

For a clean wire we can see easily that $G^0_{11}(\omega)=G^0_{22}(\omega)$ and $G^0_{12}(\omega)=G^0_{21}(\omega)$, thus the two conductivities $G^0_{31}$ and $G^0_{32}$ are equal, and by measuring the gate conductance one can extract the ideal conductance of the wire \cite{safi,hekking_blanter_buttiker}.

Moreover, if the impurity is in the middle ($x_i=0$), such that one does not break the initial mirror symmetry of the problem with respect to the origin, we find that the gate conductance is unchanged by the presence of the impurity $G_{3m}^0(\omega)=G_{3m}(\omega)$, for $m=1,2,3$.
This should be true also for any impurity distribution conserving this mirror symmetry, thus for a Gaussian extended disorder. Thus the gate offers the advantage that for a symmetric impurity distribution, one can extract directly the ideal conductance of a one-dimensional system. In a realistic experiment however, the contacts are often not perfect and can be asymmetric. In the case of asymmetric contacts the conductance of the gate is no longer dominated by the bulk impurity, but it is proportional to the asymmetry between the two contacts. This situation will be examined elsewhere.

Similarly, the non-symmetrized noise on the gate is given by:
\begin{eqnarray}
S_3(\omega) =\int_{-\infty}^{\infty} dt e^{i\omega t}
 \left\langle \delta j_3(0) \delta j_3(t)
\right\rangle =\int_{-\infty}^{\infty} dt e^{i\omega t}
 \left\langle [\delta j_2(0)+\delta j_1(0)][\delta j_2(t)+\delta j_1(t)]
\right\rangle \;.
\end{eqnarray}

It leads to:
\begin{eqnarray}
S_3(\omega) =S_{11}(\omega)+S_{22}(\omega)+S_{12}(\omega)+S_{21}(\omega)~.
\end{eqnarray}

We can also define a non-symmetrized excess gate noise as:
\begin{eqnarray}
\Delta S_3(\omega) =S_3(\omega)-S_3(\omega)|_{V=0}~.
\end{eqnarray}
We find also that, when the impurity lies exactly at the center of the wire, $\Delta S_{11}(\omega)=\Delta S_{22}(\omega)=-\Delta S_{12}(\omega)=-\Delta S_{21}(\omega)$, and the excess gate noise cancels.
The total noise in this situation is thus given by the FDT: $S_{33}(\omega)=-2 \hbar \omega N(\omega) {\rm Re}[G_{33}^0(\omega)]$.

\subsection{Non-symmetrized noise in a nanotube}
\begin{figure}[htbp]
\begin{center}
\includegraphics[width=7.5cm]{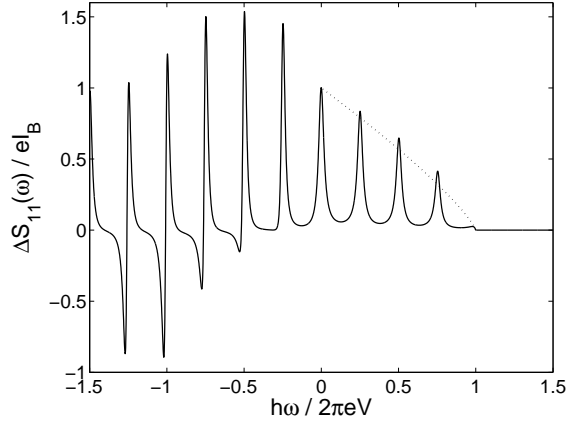}
\vspace{0.15in} \caption{\small Non-symmetrized excess noise $\Delta S_{11}$ for a nanotube divided by
$eI_B$ for the interactions parameter $g=0.25$ and for $x_i=0$, $T=0$, $\lambda/eV=0.01$ and $g\hbar\omega_L/eV=0.01$. The dotted line is the emission noise of an infinite wire with $g=g^*\approx 0.8$.}\label{nanotube}
\end{center}
\end{figure}

The analysis in the previous section was appropriate for a quantum wire with a single channel of conduction. However, realistic one-dimensional conductors such as carbon nanotubes, for which measurements of the zero frequency current-current fluctuations are now available \cite{noise0,kontos} have more channels of conduction. For example, a carbon nanotube has four channels of conduction, out of which one with an effective interaction parameter   $g \approx 0.25 e$ \cite{balentsegger}. If the impurity is in the middle of the wire, i.e. $x_i=0$, we can see that the period of the noise oscillations depends only on the fractional charge $g$ of the charge sector, and is given by $2 \pi v_F/g L$. In the limit where the tube is not too long, a slight asymmetry between the excess emission noise and excess absorption noise will start developing, but this asymmetry will not be as pronounced as in the case of a single-channel quantum wire, due to the existence of the four channels of conduction.
In the long-tube limit the effect of the extra channels of conduction will be  visible in the form of the envelope of the oscillations, where the value of $g$ which determines the exponent of the power-law dependence  will be renormalized to $g^*=(g+3)/4\approx 0.8$. This limit is presented in Fig. \ref{nanotube}. On the other hand, the averaging over the first half period of the oscillations will retrieve solely the value of the fractional charge of the charge mode $g=0.25$.

\subsection{Experimental relevance}

We should now make some comments on the accessibility of the two regimes discussed above in an experiment. For a nanotube of a micron length for example, $\pi \omega_L=v_F/g L \approx 10THz$. This corresponds to $\hbar \omega_L/k_B \approx 20K$. Thus, regime $A$, as specified in Fig.\ref{zero_frequency}, is achieved for $T \ll 20K$, and $V \ll 2 meV$, with $k_B T \ll e V$, thus temperatures of order of $0.1K$, and voltages of order of $0.1 mV$ would be appropriate. The temperature and voltage can be higher if the tube is shorter, as it is the case for example in Ref.~\cite{kontos}, where $\omega_L \approx 15meV$, and the linear regime occurs for $V$ of order of $mV$. In the frequency dependence of the noise the Josephson frequency would appear at frequencies of the order of $THz$.

On the other hand, the regime $B$ described in Fig.~\ref{excessnoise2}, occurs for the same range of temperatures, but for larger voltages, for example for the case discussed above $V \approx 10-100meV$, depending on the length of the wire. In this regime the Josephson singularity occurs for frequencies of $\approx 50 THz$. This range of frequencies is much harder to achieve experimentally, also the high voltage required will contribute to the heating of the sample. While the Josephson frequency is very high, the oscillations should be however visible for frequencies of the order of $10 THz$. This is thus the necessary frequency to achieve experimentally in order to retrieve the value of the fractional charge by performing an average of the finite-frequency noise.
We should note however that the signature of interactions is present in the AC  differential conductance and noise even at lower frequencies.


\section{Conclusions}
In this paper we have performed the first study of the differential AC conductance and the finite-frequency noise in a quantum wire connected to metallic leads in the presence of a single impurity. The single impurity scenario may correspond to either a bulk impurity, or to an impurity located at one of the contacts. While in general nanotubes are clean, and most of the backscattering comes from the imperfect contacts, the situation of a single central impurity can be achieved experimentally using for example an unbiased STM tip. In this case the effect of the bulk impurity dominates over the effect of the impurities at the contacts. Here we review some of the main results, which are also presented in more detail in the introduction.

We have found that even in the presence of leads, many signatures of interactions are still present in the behavior of the AC differential conductance and the noise, and could be observed experimentally. We have first focused on the excess AC conductance, which is defined as the difference betwen the corresponding values at finite and zero voltage, and which is proportional to the impurity strength. We have found that, while being zero for a linear system (in the absence of interactions), the excess AC differential conductance has a rich non-linear behavior for an interacting non-chiral LL. Another important observation that we made was the strong asymmetry in the finite-frequency excess noise: the emission and the absorption components of the excess noise, while identical in the absence of interactions, are different if interactions are present in the wire. We explained this asymmetry by the non-linearity in the system, showing that the difference between the emission and the absorption noise is given by the real part of the excess differential AC conductance of the wire. 

We have also established exactly a few other interesting facts about the non-symmetrized which showed the value of studying this quantity experimentally, instead of the symmetrized noise. For instance, the emission noise was shown to be equal to the partition noise in the quantum regime. 
By carrying on a perturbative analysis to lowest order in the impurity strength, we have shown that at low temperature the noise exhibits oscillations whose periodicity is inversely proportional to the value of the fractional charge. The existence of the oscillations is a crucial difference between the LL model and an alternative model, the dynamical Coulomb blockade model.  When the length of the tube is much larger than the inverse of the applied voltage, the envelope of the oscillations is given by the form of the non-symmetrized noise for an infinite LL with the same interaction parameter.
We have found that an average over the first half-period of oscillation in the long-tube limit gives directly access to the value of the fractional charge $g$ with more accuracy and less restrictions than the analysis of the symmetrized noise.

We have discussed also the presence of a gate, and have shown that for any disorder configuration with mirror symmetry the gate AC conductance is not affected by disorder, but is equal to the conductance of the clean wire, which could give access to the interaction parameter.

We have analyzed how our results change in the presence of multiple channels, such as it is the case for a carbon nanotube. Last but not least we have discussed the experimentally relevant values of the parameters in our analysis.

\section{Acknowledgments} We would like to thank C.~Glattli, H.~Bouchiat, R.~Deblock, T.~Kontos, B.~Reulet and E. Sukhorukov for helpful discussions. C.~Bena acknowledges the support of a Marie Curie Action under FP6.


\vspace{1cm}

{\noindent \Large{\bf Appendix}}
\appendix


\section{Functions $G_B$ and $S_B$}

The function $G_B$ is defined by
\begin{equation}
G_B(\omega) =\frac{1}{\hbar \omega} \int_0^\infty dt \left( e^{i \omega t}-1 \right)
\left\langle \left[ j_B(t),j_B(0) \right]
\right\rangle \; . \label{s1_C}
\end{equation}
while $S_B$ is defined as
\begin{equation}
S_B(\omega) \equiv \int_{-\infty}^\infty dt \, e^{i \omega t}
\left\langle  \delta j_B(0) \delta j_B(t)
\right\rangle , \label{s1_A}
\end{equation}
where $j_B$ is the backscattering
current operator at the backscattering site $x_i$ defined in Eq.~(\ref{jb1_def}).

The evaluation of $G_B$ and $S_B$ perturbatively up to second order in $\lambda$ gives \cite{bena}:
\begin{eqnarray}
G_B(\omega)&=&\frac{1}{2\hbar \omega}\frac{e^2\lambda^2}{\hbar^2} \int_0^\infty dt (e^{i \omega t}-1)
\cos\left(\frac{e V t}{\hbar}\right)\sum_{s=\pm}s e^{4 \pi {\cal C}(x_i,s t;x_i,0)}\, ,\nonumber\\
\end{eqnarray}
where the two-point functions ${\cal C}$ is presented in Appendix B. The sum over $s$ can be expressed as:
\begin{eqnarray}
\sum_{s=\pm}s e^{4 \pi {\cal C}(x_i,s t;x_i,0)}&=&2i\sin\{4\pi{\rm Im}[{\cal C}(x_i,t;x_i,0)]\} e^{4\pi{\rm Re}[{\cal C}(x_i,t;x_i,0)]}~.
\end{eqnarray}
Similarly we can write $S_B(\omega)=[f_A(\omega)-\hbar \omega G_B(\omega)-\hbar \omega G_B(-\omega)]/2$
where
\begin{eqnarray}
f_A(\omega)&=&
i \frac{e^2\lambda^2}{2\hbar^2} \sum_{m=\pm 1} \coth \Big[ \frac{\hbar
\omega+m e V}{2 k_B T}\Big]\nonumber\\
&\times&\int_0^\infty dt \sin[(\omega+m e V/\hbar) t] \sum_{s=\pm}s e^{4 \pi {\cal C}(x_i,s t;x_i,0)} .\nonumber\\
\end{eqnarray}

\section{Green's function ${\cal C}^{\cal R}$}

The Green's function ${\cal C}^{\cal R}$ is given by the Fourier transform:
\begin{eqnarray}
\tilde{\mathcal{C}}^{\cal R}(x,y,\omega)\equiv\int_{-\infty}^{\infty} \,
e^{i \omega t} \, {\mathcal{C}}^{\cal R}(x,t;y,0) \, dt~,
\end{eqnarray}

where,
\begin{eqnarray}
{\cal C}^{\cal R}(x,t;y,0)&=&2i\theta(t){\rm Im}[{\cal C}(x,t;y,0)]~,
\end{eqnarray}

and ${\cal C} \, = \, {\cal C}_{\rm GS}  \, + \, {\cal C}_{\rm TF}$.
The ground state (GS) and thermal fluctuations (TF) contributions are given by \cite{dolcini}:
\begin{eqnarray}\label{CGS}
\lefteqn{{\cal C}_{\rm GS}(x,t;y,0)  =}  \\
& & -\frac{g}{4 \pi} \left\{ \sum_{m \in Z_{\rm even}}
\gamma^{|m|} \ln{\left(\frac{ (a+i
\tau)^2+(\xi_r+m)^2}{a^2+ m^2 }\right) } \,
\right. \label{Creg_GS} \nonumber\\
& & \hspace{1cm} + \left. \sum_{m \in Z_{\rm odd}} \gamma^{|m|}
\left\{ \ln{\left(\frac{ (a+i
\tau)^2+(m-\xi_R)^2}{a^2+(m-\xi_R)^2 }\right) } \, + \,
\frac{1}{2} \, \ln{ \left(\frac{[a^2+(
\xi_R+m)^2]^2}{[a^2+(2 \xi+m)^2] \,[a^2+(2 \eta+m)^2]
}\right)}
\right\} \right\} \, ,  \nonumber
\end{eqnarray}

and,
\begin{eqnarray}\label{CTF}
\lefteqn{{\cal C}_{\rm TF}(x,t;y,0) = }  \\
& & -\frac{g}{4 \pi} \left[ \sum_{m \in Z_{\rm even}} \gamma^{|m|}
\sum_{r=\pm} \,\ln \left( \frac{\sinh{[\pi\Theta (\tau+r(\xi_r+m))
]}  }{\pi \Theta (\tau+r(\xi_r+m)) }  \frac{\pi \Theta m}{
\sinh{[\pi \Theta m]} } \right) \right.
 \nonumber \\
& & \displaystyle \hspace{1.0cm}+ \!\!\! \sum_{m \in Z_{\rm odd}}
\gamma^{|m|}  \sum_{r=\pm} \, \ln \left( \frac{\sinh{[\pi \Theta
(\tau+r(m-\xi_R)) ]}}{\pi \Theta(\tau+r(m-\xi_R))}  \,
 \frac{\pi \Theta (m-\xi_R)}{\sinh{[\pi \Theta  (m-\xi_R)]}}
 \right)  \nonumber  \\
& & \displaystyle \hspace{1.0cm}+ \left. \sum_{m \in Z_{\rm odd}}
\gamma^{|m|} \ln \left(\frac{\sinh^2{[\pi \Theta (\xi_R+m)
]}}{[\pi \Theta(\xi_R+m)]^2}  \frac{\pi \Theta (2 \xi + m) } {
\sinh{[\pi \Theta(2 \xi + m)]}}  \frac{\pi \Theta (2 \eta + m) } {
\sinh{[\pi \Theta(2 \eta+ m)]}}\right)
 \right]  \; ,  \nonumber
\end{eqnarray}

where $\gamma=(1-g)/(1+g)$, $\xi=x/L$, $\eta=y/L$, $\xi_r=(x-y)/L$, $\xi_R=(x+y)/L$,
$\tau=t \omega_L$, $\Theta=k_B T/
\hbar \omega_L $, and $a=\omega_L/\omega_c$ is the
(dimensionless) inverse cut-off.\\


\section{Backscattering current}

The averaged backscattering current is given by \cite{dolcini}:
\begin{eqnarray}
I_{B}=\frac{e\lambda^2}{4\hbar^2}\int_{-\infty}^{\infty} dte^{i e V/\hbar t}\sum_{s=\pm}
se^{4\pi {\cal C}(x_i,st;x_i,0)}~.
\end{eqnarray}

With the help of the parity properties of the Green's function, it can been shown that the
imaginary part of the backscattering current cancels. As a consequence, $I_{B}$ is purely real and is given by:
\begin{eqnarray}
I_{B}&=&-\frac{e\lambda^2}{2\hbar^2}\int_{-\infty}^{\infty} dt\sin\left(\frac{e V t}{\hbar}\right)\sin\{4\pi{\rm Im}
[{\cal C}(x_i,t;x_i,0)]\} e^{4\pi{\rm Re}[{\cal C}(x_i,t;x_i,0)]}~.
\label{ib}
\end{eqnarray}
The behavior of the backscattered current was analyzed in detail is Refs.~\cite{dolcini,dolcini_current}.


\begin{thebibliography}{99}
\bibitem{note}A general derivation of the positivity of $S_B$ defined in Eq.~(\ref{sb1}), and implicitely of $S_{11}$ at positive frequencies is presented in Ref.~\cite{ines_FDT}.
\end{thebibliography}
\end{document}